\documentclass[secnumarabic, twocolumn, graphics,floatfix, superscriptaddress,tightenlines, aps, prb]{revtex4-1}
\usepackage[dvips]{graphicx}
\usepackage{dcolumn}
\usepackage{bm}
\usepackage{color}

\begin{document}

\title{Nuclear spin relaxation in n-GaAs: from insulating to metallic regime }

\author{M.~Vladimirova}
\affiliation{Laboratoire Charles Coulomb, UMR 5221 CNRS-Universit\'{e}  de Montpellier,
F-34095, Montpellier, France}
\author{S.~Cronenberger}
\affiliation{Laboratoire Charles Coulomb, UMR 5221 CNRS-Universit\'{e}  de Montpellier,
F-34095, Montpellier, France}
\author{D.~Scalbert}
\affiliation{Laboratoire Charles Coulomb, UMR 5221 CNRS-Universit\'{e}  de Montpellier,
F-34095, Montpellier, France}

\author{M.~Kotur}
\affiliation{Ioffe Physico-Technical Institute of the RAS, 194021 St.Petersburg, Russia}
%

\author{R. I. Dzhioev}
\affiliation{Ioffe Physico-Technical Institute of the RAS, 194021 St.Petersburg, Russia}

\author{I.~I.~Ryzhov}
\affiliation{Spin Optics Laboratory, St. Petersburg State University, 1 Ul'anovskaya,
Peterhof, St. Petersburg 198504, Russia}

\author{{G.~G.}~Kozlov}
\affiliation{Spin Optics Laboratory, St. Petersburg State University, 1 Ul'anovskaya,
Peterhof, St. Petersburg 198504, Russia}

\author{V.~S.~Zapasskii}
\affiliation{Spin Optics Laboratory, St. Petersburg State University, 1 Ul'anovskaya,
Peterhof, St. Petersburg 198504, Russia}

\author{A. Lema\^{\i}tre}
\affiliation{Centre de Nanosciences et de nanotechnologies - CNRS - Universit\'{e} Paris-Saclay - Universit\'{e} Paris-Sud, Route de Nozay, 91460 Marcoussis, France}

\author{K.~V.~Kavokin}
\affiliation{Spin Optics Laboratory, St. Petersburg State University, 1 Ul'anovskaya,
Peterhof, St. Petersburg 198504, Russia}
\affiliation{Ioffe Physico-Technical Institute of the RAS, 194021 St.Petersburg, Russia}

\begin{abstract}
Nuclear spin relaxation is studied in n-GaAs thick layers and microcavity samples with different electron densities.
We reveal that both in metallic samples where electrons are free and mobile, and in insulating samples, where electrons are localized, nuclear spin relaxation is strongly enhanced at low magnetic field.
The origin of this effect could reside in the quadrupole interaction between nuclei and fluctuating electron charges, that has been proposed to drive nuclear spin dynamics at low magnetic fields in the insulating samples.
The characteristic values of these  magnetic fields are given by dipole-dipole interaction between nuclei in bulk samples, and are  greatly enhanced in microcavities, presumably due to additional strain, inherent to micro and nanostructures.
\end{abstract}

 \maketitle

\section{Introduction}
\label{sec:intro}
Magnetic field and density dependence of the electron spin relaxation in n-doped semiconductors has been extensively studied during past decades \cite{DyakonovBook}.  It is well established that, at low temperatures, spin relaxation of electrons in lightly doped bulk semiconductors,
like GaAs, and nanostructures (quantum wells and quantum dots) is determined by the contact hyperfine
interaction with lattice nuclei \cite{Dyakonov:1974, Merkulov2002}.  The electron spin, hopping over shallow donors, feels
a fluctuating nuclear magnetic field, which makes its spin flipping. In GaAs, at impurity concentrations
 $n\approx 10^{14}-10^{15}$~cm$^{-3} $ the nuclear field is dynamically averaged, because the typical hopping time   ($1- 0.01$ ns)
 is much shorter than the average period of electron spin precession in the random nuclear field \cite{OpticalOrientation}.
 With doping, the hopping rate $1/\tau_c$ increases exponentially. \cite{Dzhioev2002}
 As a result of more effective averaging of
 random nuclear fields, the nuclear-mediated electron spin relaxation time,  $T_s$, becomes longer, so that  another relaxation mechanism, based on spin-orbit interaction, takes over.\cite{Kavokin2001,KavokinSST09,Dzhioev2002}. When the concentration of donors is further increased above the metal-to-insulator transition (MIT),  which occurs in GaAs at $n_D= 2 \times 10^{16}$~cm$^{-3}$, the electron spin relaxation time start to decrease, because in the metallic phase, the Dyakonov-Perel (DP) mechanism
  dominates spin relaxation of the Fermi-edge electrons \cite{OpticalOrientation}.
  Thus, the density dependence of the electron spin relaxation time in n-doped semiconductors  is strongly non-monotonous \cite{Dzhioev2002, Sprinzl2010}.

 Much less is known about nuclear spin-lattice relaxation times, $T_1$.
 Most of the studies were carried out in the presence of the external magnetic field stronger than local field $B_L$, that characterizes dipole-dipole interactions between nuclei $B>>B_L\sim 2$~G in GaAs \cite{Kikkawa2000,Lu,King2012,DyakonovBook}.
 Moreover, in most of the existing optical detection protocols, it is necessary to inject out-of-equilibrium carriers, in order to probe any changes in the electron spin polarization or splitting, induced by nuclear spin \cite{Paget82,Urbaszek}.
Because injected electron spins are not in thermal equilibrium, they strongly affect nuclear spin dynamics.
This leads to various nonlinearities \cite{Urbaszek, OpticalOrientation}, and makes it difficult to address nuclear spin relaxation.

 Let us summarize  what is known about nuclear spin-lattice relaxation times in n-GaAs.
 At low temperatures the 
 nuclear spin relaxation in n-doped semiconductors is mediated by electrons \cite{Abragam}.
  The relaxation of the nuclear spins situated under the donor orbits is rather fast (fraction of a second), while the relaxation of the remote nuclei can be much slower, because it is dominated by spin diffusion towards the donors \cite{Paget82,Giri2013}.
The characteristic time for this diffusion-limited
relaxation can be estimated as \cite{DeGenes}:
\begin{equation}
T_D^{-1}\approx 4 \pi D n_d a,
\label{eq:Td}
\end{equation}
where $D \approx
10^{-13}$~cm$^2/$s is the nuclear spin diffusion coefficient
\cite{Paget82},  $a=10$~nm is the localization radius of the electron
on the center, and $n_d$ is the donor density. It can reach minutes and even hours in very dilute samples, but shortens close to MIT.

With increasing the donor density above MIT, nuclear relaxation is expected to slow down, because electrons are no more localized on the donor sites. It is no more limited by diffusion, but by the spin fluctuations of  free Fermi-edge electrons (Korringa mechanism) \cite{Korringa1950,Lu,KoturJETP2014}.
However, in the intermediate regime, close to MIT, pairs of closely spaced donors still act as localizing centers for electrons. They were shown to contribute to the nuclear spin relaxation. As a result, it is still limited by  spin diffusion, rather than by the Korringa mechanism.\cite{Giri2013}

This qualitative picture  based on these three relaxation mechanisms (hyperfine interaction, nuclear spin diffusion and Korringa mechanism)  describes reasonably the existing experimental data at low temperatures and strong fields.
But it does not predict any substantial modification in the low field regime $B<B_L$, a peculiar regime where 
one must distinguish the longitudinal nuclear spin relaxation time from the time characterizing the warm up of the nuclear spin system, which is determined by energy transfer between the nuclear spin and the crystal lattice \cite{Abragam,OpticalOrientation}. 
Indeed, these relaxation times coincide only at magnetic fields much larger than the local field, while at low magnetic field the relaxation time of the non-equilibrium nuclear spin  becomes much shorter due to dipole-dipole nuclear interactions.
The characteristic time of the dipole-dipole relaxation, also referred to as the transverse spin relaxation time, is as short as 
$T_2 \sim 100$~$\mu$s \cite{OpticalOrientation}.
Therefore, any low-field nuclear spin polarization, showing relaxation times longer than $T_2$, is in fact a quasi-equilibrium polarization.
Its value is uniquely defined by the applied magnetic field and the 
nuclear spin temperature $\Theta_N$.
The concept of the spin temperature is therefore essential for the description of the low-field nuclear spin dynamics \cite{AbragamProctor}.
The relaxation time of the nuclear polarization in this regime is given by the  relaxation of the nuclear spin temperature to the lattice temperature. 
This process is often referred to as a warm up of the nuclear spin system \cite{OpticalOrientation}. 
We note, that the low-field regime is particularly important when deep cooling of  the nuclear spin system
is intended,  because the demagnetisation to low field is required in this protocol \cite{KALEVICH1982}.

%
We have recently reported a strong enhancement of the nuclear spin warm-up rate in a n-GaAs bulk sample in the insulating regime\cite{Kotur2016}. This surprising effect could be understood  by taking into account an additional relaxation mechanism:  the interaction of nuclear quadrupole moments with electric field gradients induced by slow spatiotemporal fluctuations of localized electron charges.

In this paper, we scrutinize  nuclear spin dynamics in six n-GaAs samples  with the concentration varying across MIT from $2\times10^{15}$~cm$^{-3}$ to $9\times10^{16}$~cm$^{-3}$.
Our goal is to provide a comprehensive picture of  (i)  the spin relaxation efficiency of the bulk nuclei, situated outside of the donor-bound electron Bohr radius $a_B=10$~nm, and (ii)  magnetic field dependence of nuclear spin relaxation in samples with different donor densities.
The experiments reported in this paper involve three different experimental techniques, all using different multi-stage strategies, in order to separate preparation of nuclear spin under optical pumping from the the measurements of the spin relaxation: photoluminescence  (PL) with dark intervals \cite{KALEVICH1982,KoturJETP2014,Kotur2016}, Faraday rotation (FR) \cite{Giri2013}, and spin noise (SN) spectroscopy \cite{Ryzhov2015,Ryzhov2016}.
This choice of the methods allows  for the comparison between bulk GaAs layers and microstructures, thin layers embedded in planar microcavities, that were used to amplify SN and FR signal induced by nuclear spin polarization \cite{Kavokin1997,Salis2005,Cherbunin2015}.
The main results of our analysis can be summarized in three points.
(i) At strong magnetic fields the spin relaxation rate
fits reasonably the picture described in the Introduction, based on the hyperfine interaction, nuclear spin diffusion and Korringa mechanism. In this study, strong magnetic fields designate fields much larger than the local field but not exceeding 1000~G.
%
%
(ii) At low magnetic fields,  quadrupole-induced  enhancement of the warm up rate appears to be ubiquitous, it shows up  in all the samples.
(iii) The characteristic field $B_{1/2}$ that controls the onset of the nuclear spin warm up enhancement is of order of the local field $B_L$
in bulk samples, but is up to $6$ times higher in all microstructures, either sandwiched between the Bragg mirrors, or between two GaAlAs barriers.
%
 We attribute this difference to the small, but not negligible strain present in all microstructures, and the resulting quadrupole splittings between nuclear spin states.

The  paper is organized as follows. In Section \ref{sec:samples}, we describe the samples studied in this work. In Section \ref{protocols}, we present three different types of experiments used for studies of nuclear spin relaxation, and the procedure applied to extract bulk nuclear spin relaxation times. In Section \ref{discussion} we present the results of the measurements, and draw up the picture of nuclear spin relaxation in n-GaAs: magnetic field, temperature and donor density dependence, as well as the effect of  microstructures on the
nuclear spin relaxation.
The experimental results are compared with the existing models for nuclear spin relaxation, that allow us to partly understand the data.
Possible explanations for
the enhancement of the nuclear spin relaxation rate at low magnetic field and the role of the microstructures in this
phenomenon are also discussed.
The results of the work are summarized in Section \ref{Conclusions}.

\section{Samples}
\label{sec:samples}
We use in this work six different Si-doped GaAs samples.
Two    GaAs
layers with Si donor concentration of $n_d=4 \times10^{15}$ cm$^{-3}$ (Sample D) and
$n_d=6 \times10^{15}$ cm$^{-3}$ (Sample C) were grown on $500$~nm-thick GaAs substrates by liquid (Sample D) or gas (Sample C) phase epitaxy.
The thicknesses of these layers are $20$ $\mu$m (Sample D) and $200$ $\mu$m (Sample C).
These epitaxial layers are  so thick, that we will refer to Samples D and C as bulk samples.
%
%

Three microcavity samples were grown by molecular beam epitaxy.
In these structures, a Si-doped $3\lambda/2$ GaAs cavity layer is
sandwiched between two Bragg mirrors, in order to enhance the sensitivity of Faraday rotation and spin noise experiments.
%
The front (back) mirrors are distributed Bragg reflectors composed
of $25$ ($30$) pairs of AlAs/Al$_{0.1}$Ga$_{0.9}$As layers, grown on
a $400~\mu$m thick GaAs substrate.
Due to multiple reflections from the mirrors, the FR ( SN) is amplified  by a
factor $N\sim 1000$ with respect to the bare  cavity layer, corresponding to the interaction length
$L=0.7$~mm.
The  cavity was wedge shaped in order to have the possibility to tune the cavity mode energy by varying the spot position on the sample.
 The detuning between the energy gap of undoped GaAs, chosen as a reference, and the cavity mode could be slightly varied.
 Here we worked typically around $20-30$~meV, depending on the sample.
Note, that  because of this large  detuning of the cavity mode with respect to the GaAs energy gap, the interband emission is strongly suppressed. Therefore, studies of nuclear spin dynamics via the degree of circular polarization of photoluminescence were not possible in the microcavity samples.
The  concentrations of Si donors were $n_d=4\times 10^{16}$~cm$^{-3}$
 (metallic, Sample A),  $n_d=2\times 10^{16}$~cm$^{-3}$
 (close to metal-insulator transition, Sample B), and $n_d=2\times 10^{15}$~cm$^{-3}$ (insulating, Sample C).

 The last Sample F was also grown by molecular beam epitaxy on GaAs substrate.
 It  is a $1$ $\mu$m-thick layer of GaAs with donor concentration $n_d=9\times 10^{16}$~cm$^{-3}$, sandwiched  between AlGaAs barriers.
 Because of the small thickness of the layer, similar to that of the microcavity samples, we will refer to this sample as a microstructure, rather
 than a bulk layer, in contrast with Samples C, D.
Thus, we have three metallic (A, B, F) and three insulating (C, D, E) samples,  among which two are bulk thick layers (C, D), and four others are  various microstructures. All  the samples have been studied in our previous works. \cite{Giri2012,Giri2013, Ryzhov2015,Dzhioev2002,Kotur2016}
 %
 %
%
%
\begin{table}[h]
\begin{center}
\begin{tabular}{|c|c|c|c|c|c|c|}
\hline
Sample  & A  & B & C & D & E & F \\ \hline
$n_d$ ($10^{15}$ cm$^{-3}$) & \textbf{$40$} & \textbf{$20$} & \textbf{$6$} & \textbf{$4$}  & \textbf{$2$} & \textbf{$90$} \\ \hline
$T_s$ (ns) & \textbf{$30$} & \textbf{$250$} & \textbf{$120$} & \textbf{$180$}  & \textbf{$80$} & \textbf{$20$} \\ \hline
layer thickness ($\mu$m)   & {0.37} & {0.37} & {200}      & {20}     & {0.37} & {1}   \\ \hline
cavity (yes/no)                                                   & {yes} & {yes} & {no}      & {no}     & {yes} & {no}   \\ \hline
measurements                                                 & {SN} & {FR} & {PL}      & {SN, PL}     & {FR, SN}  & {PL}   \\ \hline
\end{tabular}
\caption{Sample parameters: electron density, electron spin relaxation time, layer thickness, the presence of the cavity, and the type of experiments that were realized are indicated.}
\label{tab:samples}
\end{center}
\end{table}

\section{Experimental protocols}
\label{protocols}
All the experiments are realized at cryogenic temperatures, with the possibility to apply magnetic field in an arbitrary configuration.
The geomagnetic field is compensated with the precision of at least $\approx 0.1$~G.
The three types of experiments exploited in this work aim at measuring nuclear spin relaxation dynamics as a function
of the external magnetic field but in the absence of the optically created charge carriers.
Thus, the experimental protocols that we used for these studies have an important common point.
Namely, nuclear spin cooling is separated in time from the measurement stage. Cooling is always achieved via optical pumping of
the resident electrons, which, in the presence of the magnetic field component parallel to the light, is accompanied by dynamic polarization of nuclei. Then, nuclear spin relaxation in the absence of optical pumping is studied under arbitrary magnetic field.
The details of the experimental protocols that we adopted are presented below,  illustrated by  typical measurements of PL, FR and SN in our samples.
Although we have already presented each of this techniques separately  in our previous publications, we give an overview of all of them, for the sake of completeness.
%
%
%

\subsection{Photoluminescence measurements}
The experimental setup used for PL experiments  is shown in Fig.~\ref{setup}~(a).
The excitation beam was provided by
a Ti-sapphire laser at $E=1.55$~eV, circularly polarized and focused on $50$~$\mu$m-diameter spot on the sample surface.
The PL was collected in the reflection geometry, passed through a circular polarization analyzer (consisting of a photoelastic modulator (PEM) and a linear polarizer) and spectrally dispersed with a double-grating spectrometer. The signal was detected by an avalanche photodiode, connected to a two-channel photon counter synchronized with the PEM.
 External magnetic field  $B$ was applied in the oblique but nearly Voigt geometry ($<10 $ degrees), in order to allow for both dynamic nuclear polarization (here the longitudinal component $B_z$ of the applied magnetic field is important) and detection of the nuclear polarisation via the Hanle effect (here the in-plane component $B_x$ of the applied magnetic field is required).
A typical PL measurement in the  bulk GaAs sample (Sample D) is presented in Fig.~\ref{protocol}~(b).
During the pumping stage, the magnetic field $B_{pump}$
and the pumping beam are switched on.
The magnitude of the magnetic field was chosen to ensure the best sensitivity of the PL polarization to the nuclear field (close to the half width at half maximum (HWHM) of the Hanle curve).  
The nuclear spin polarization builds up on the scale of several minutes.
The duration of this stage is fixed to $T_{pump}=5$~min. After that, the pump beam is switched off,
and the magnetic field is set to the value $B_{dark}$ at which we want to study the relaxation of nuclear spin polarization.
The second stage of PL experiment will be referred to as the dark stage, its duration $T_{dark}$ was varied, in order to access nuclear relaxation dynamics.
Immediately (on the scale of electron spin relaxation time, $\tau_s$) after switching off the pump, electron spin
polarization returns to its equilibrium value (close to zero in our experimental conditions). %

Nuclear spin relaxation time $T_1$ is much longer than that of electrons.
During $T_{dark}$ the nuclear polarization (or equivalently, inverse spin temperature)
decreases by the factor of  $\exp({-T_{dark}/T_1})$.
Because the PL signal is strictly zero during this stage, it is impossible to monitor in real time the evolution of the nuclear spin polarization.
The value of the Overhauser field $B_N$ achieved after $T_{dark}$ is measured during the third stage of the PL measurement protocol.
To do so, the light is switched back on, measuring field  $B_{pump}$ is restored and the  degree of
circular polarization of PL is detected.
%
%
Measuring the  PL polarization degree in the
beginning of the third stage $\rho_{dark}$, as a function of $T_{dark}$ provides the information
on the relaxation of nuclear field for a given value of the magnetic field applied during second (dark) stage.
To increase the precision of $\rho_{dark}$ measurements,  we monitored the PL polarization exponentially approaching its equilibrium value during $150$~s.
The Overhauser field achieved after the dark stage $B_N(T_{dark})$ is related to PL polarization $\rho_{dark}$. It is given by the Hanle formula:
\begin{equation}
B_N(T_{dark})=B_{1/2} \sqrt { \frac {\rho_0-\rho_{dark} }{\rho_{dark}}}-B_{pump},
\end{equation}
where $\rho_0$ is the PL polarization in the absence of the external field, and $B_{1/2}$ is the half width of the Hanle curve, measured independently \cite{KoturJETP2014} under conditions where nuclear spin polarization is absent (pump polarization modulated with PEM at $50$~kHz).
For the shortest $T_{dark}$ we have checked that the Overhauser field $B_N(T_{dark})$ 
restores to its  value measured  at the end of the pumping stage.
%
%
Thus, measuring $B_N(T_{dark})$ as a function of $T_{dark}$ and fitting the resulting exponential decay with the function
$B_N(T_{dark})=\exp{(-T_{dark}/T_1)}$ we can access the nuclear spin relaxation time $T_1$ for a given external magnetic field applied during the dark stage.
Note, that  a similar protocol has been first proposed and realized by Kalevich et al. \cite{KALEVICH1982},  and then further developed in Refs. \onlinecite{KoturJETP2014, Kotur2016}.
%
%
\subsection{Faraday rotation measurements}
\begin{figure}[t!]
\center{\includegraphics[width=0.8\linewidth]{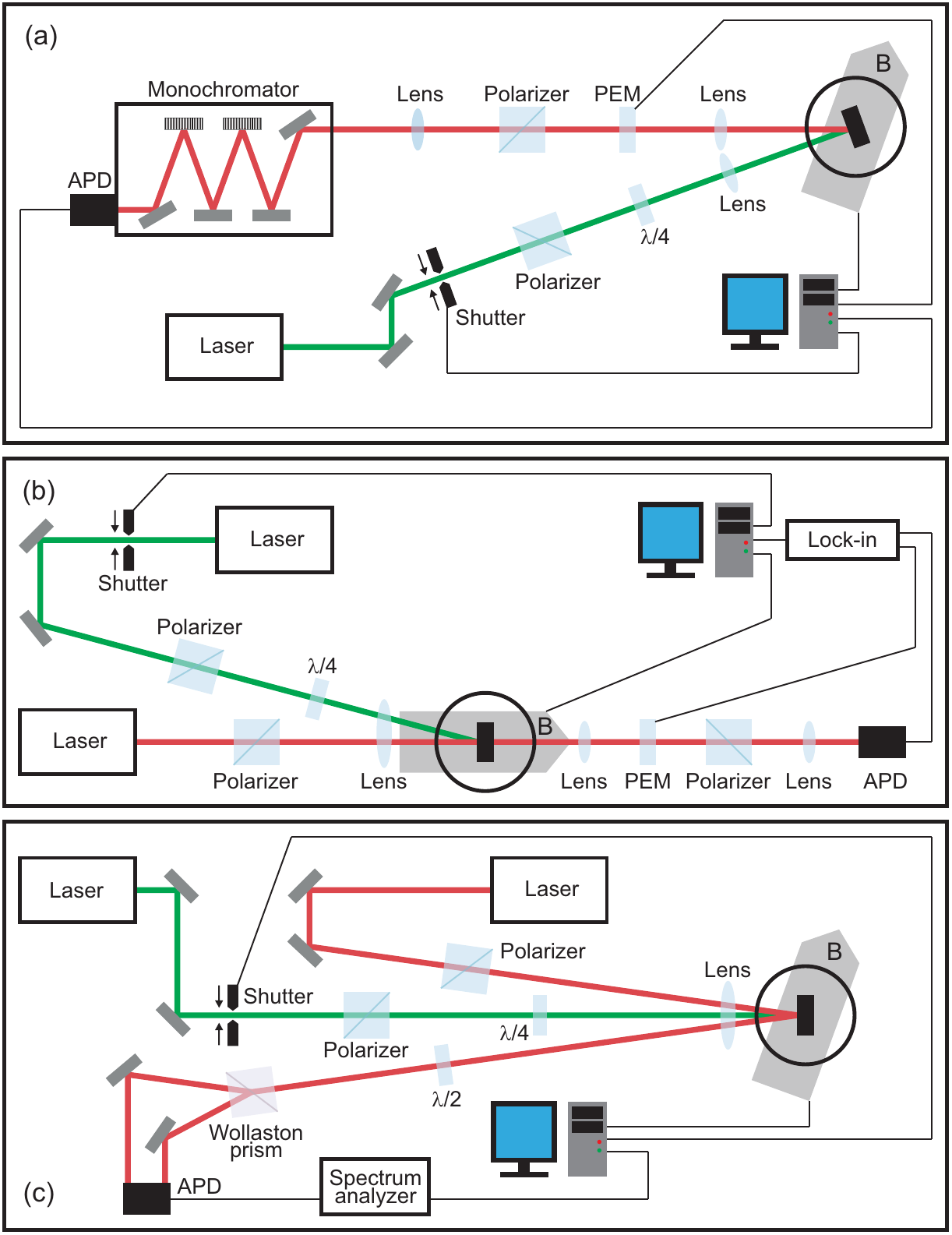} } \caption{
Sketch of experimental setup used for PL (a), FR  (b) and SN (c)  measurements } \label{setup}
\end{figure}

The  experimental protocol that we use in this work was initially proposed in Ref. \onlinecite{Artemova1985}, and then successfully
implemented in Ref. \onlinecite{Giri2013}.
FR setup is shown in Fig.~\ref{setup}~(b).
For the microcavity samples it is not possible to work under resonant pumping conditions, because the band
edge is situated in the middle of the Bragg stop-band. \cite{Giri2012}
Therefore, we used a cw-laser diode emitting at $1.59$~eV, well above the Bragg stop-band, for both FR and SN measurements.
Here we worked  in the transmission geometry, and purely longitudinal magnetic field $B_z$ was applied (parallel to the growth axis).
In contrast with PL, it is  a two-beam experiment.
While the pump beam
was focused into a $0.5$~mm diameter spot on the sample surface, the linearly polarized probe beam was focused
on a $50$~$\mu$m spot,  to probe selectively the area homogeneously excited by the pump beam.
A linearly polarized probe beam was provided by a mode-locked Ti-sapphire laser, to ensure better stability of the transmission through the cavity mode. It was spectrally filtered with the 4f zero-dispersion line,
down to the spectral width of $5$~meV.
Its energy was fixed at the cavity mode, which filters the incident pulse at the cavity mode energy, corresponding to the
detuning of $\approx 20-30$~meV, with respect to the GaAs band gap. Typical pump and probe powers are $10$ and $2$~mW, respectively.
The rotation of probe polarization after transmission from the sample was analysed by PEM operating at $100$~kHz,
followed by the linear polarizer, the resulting signal was sent into avalanche photodiode and demodulated at
PEM frequency. Both metallic (Sample B) and insulating  (sample E) samples were studied by the  FR technique.

A typical FR measurement for Sample E is presented in Fig.~\ref{protocol}~(b).
A nonzero FR is always measured in  presence of the external magnetic field.
This static field-induced FR is not related to the photoinduced  polarization of electrons and nuclei and was
systematically subtracted from all the measurements.
The pumping stage of the experiments starts when both pump beam and pumping
magnetic field $B_{pump}=180$~G are switched on.
During pumping, FR is continuously monitored. One can see,
that the signal increases with roughly two characteristic times.
We systematically observe a fast increase on the scale of several seconds, followed by a slower growth on minutes scale. Full saturation is eventually achieved after $\approx 1$~hour for Sample E, corresponding to the maximum Overhauser field achievable under given pumping conditions.
To explain this dynamics, we argue that FR signal under optical pumping consists of two contributions.
The strongest signal comes from spin-polarized electrons, which are bound to donors in the insulating Sample E, and experience the effect of the increasing Overhauser field.
This component grows with two characteristic times.
 Fast increase is determined by the polarization of nuclei close to donors within
bound electron Bohr radius.
%
Obviously, in metallic samples  fast component is absent, because nuclei
are rather homogeneously polarized by Fermi-edge electrons \cite{Giri2013}.
The second contribution comes from conduction band spin splitting induced
by the Overhauser field averaged over the measured sample volume.
Its amplitude is small with respect to the first component, and the dynamics  develops on the nuclear spin diffusion time-scale.
Although the observed dynamics is quite interesting by itself, this work is focused on nuclear spin dynamics in the absence of  optical pumping and we will not consider the dynamics during pumping further.

Thus, we concentrate on the second stage of FR experiments.
After the time $T_{pump}=3$~minutes that we keep fixed in these experiments, the pump beam is switched off, and the
magnetic field is set to $B_{dark}$,  for which we want to study the relaxation of nuclear spin polarization.
All photoinduced electron spin polarization relaxes on the scale of $T_s$, which is not resolved in these experiments,
and the remaining  FR signal is small (Fig.~\ref{protocol}~(c)).
It is almost exclusively determined by electron spin band splitting induced
by Overhauser field averaged over the measured sample volume.
This is the dominant mechanism of FR in the absence of optical pumping.~\footnote{For Sample B,
with electron density close to MIT, the contribution due  to spin polarization of electrons bound to donors can also be detected. \cite{Giri2013}
}
Because FR is directly proportional to the average Overhauser field in the probed volume,
$\Theta_F=V_N B_N L_{eff}$, where  $L_{eff}$ is the effective length of the cavity.
 The effective length of the cavity is the double thickness of the cavity $2L$ times the number
 of round trips $L_{eff}=2 L\times N=Q \lambda_{res}/(2 \pi)$, where $Q=19280\pm480$ is the quality factor measured by interferometrique technique, $\lambda_{res}$ is the wavelength of the  cavity resonance.\cite{Giri2013}
 Therefore, for our $3 \lambda_{res}/2$ cavity we get $L_{eff}\sim 0.7$~mm.
 Because $L_{eff}>>2 L$, Faraday rotation is strongly enhanced with respect to
 a bare layer with thickness $L$.\cite{Kavokin1997,Salis2005,Cherbunin2015}
The proportionality coefficient $V_N$ is called nuclear Verdet constant,
 in analogy with traditional Verdet constant which characterizes the efficiency  of traditional Faraday rotation.
 The determination  of $V_N$ for each sample requires careful measurements of $B_N$ from independent experiments \cite{Giri2012}.
Nevertheless,  by  fitting the observed exponential decay of FR during the dark stage, we recover $T_1$, the relaxation time for nuclear spins in the absence of optical pumping and under arbitrary magnetic field $B_{dark}$.
In Sample B, where the decay of the nuclear polarization is additionally contributed by the  initial fast decay due to electron localisation on the donor pairs \cite{Giri2013}, we only keep the slow component of the decay, associated with the bulk nuclei.
\subsection{Spin noise measurements}
The SN setup is shown in Fig.~\ref{setup}~(c).  As for FR, we use two laser beams, one for dynamic nuclear polarization, the other for detection of the resulting nuclear spin dynamics.
Optical pumping at $1.59$~eV  with circularly polarized beam is achieved using the same laser diode.
In the same manner as in PL experiments, the magnetic field is applied  in oblique but nearly Voigt geometry (at $15$ degrees), 
in order to allow for both optical pumping of the nuclear spin and for detection of the resulting nuclear field via the peak frequency shift in the spin noise spectrum.
A linearly polarized probe beam resonant with the cavity mode is provided either by a continuous wave Ti-Sapphire laser or by a tunable external-cavity diode laser, with typical power $P_{pr} =0.25$~mW, and focused on $30$~$\mu$m.
The electrons spin fluctuations produce fluctuations of the Faraday rotation angle of the probe beam reflected from the sample, which are detected by means of a polarization sensitive optical setup, with a detector bandwidth up to 1 GHz. The SN spectra are then obtained by feeding the signal into a Fourrier transformation-based  spectrum analyzer.
%
The position of the SN peak in the spectrum is determined by the magnitude of the total magnetic field,
acting upon the electron spins.
It is given by the sum of external field $B$, and the Overhauser field $B_N$.
The conduction band electron gyromagnetic ratio for GaAs  $\gamma_e=0.64$~MHz/G is well known.
 This allows  us to directly relate the  frequency of the peak $\nu$ measured in SN spectrum to $B_N$,
 for a given arbitrary value of the applied field:
 \begin{equation}
 B_N=(\nu-\gamma_e B)/\gamma_e
 \end{equation}
The accumulation time of a SN spectrum could be reduced down to $1.5$~s without affecting measurement accuracy.
This defines temporal resolution of the experiments.
Thus, measuring time evolution of the SN peak frequency allows for determination of the nuclear spin relaxation times under
arbitrary magnetic field.

\begin{figure}[t!]
\center{\includegraphics[width=0.8\linewidth]{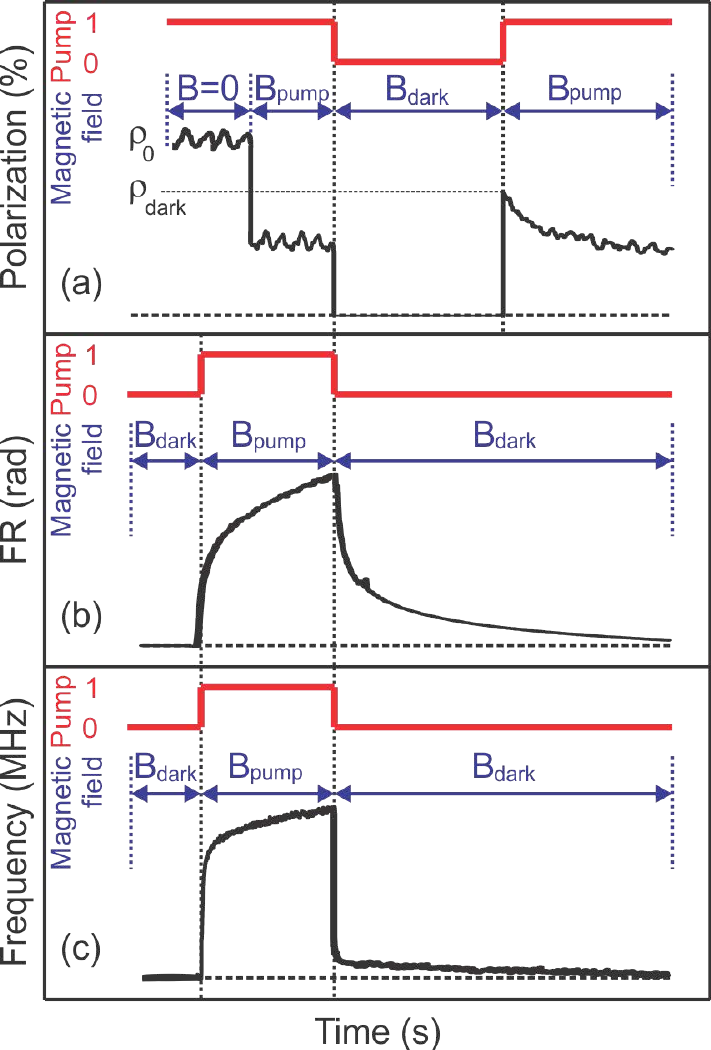} } \caption{
Typical examples of raw measurements using  three different experimental techniques: PL (a), FR (b) and SN (c). Red lines indicate the pumping slots, blue arrows indicate when magnetic field is changed, black lines are the data.
} \label{protocol}
\end{figure}

To access nuclear spin relaxation by SN measurements we adapted the procedure similar to Ref. \onlinecite{Ryzhov2015}.
It is illustrated for Sample E in Fig.~\ref{protocol}~(c), where the position of the SN spectral peak is shown as a function of time.
Before the beginning of the experiment,  nuclear spin polarization is zero.
The position of the peak in the SN spectrum is given by $\gamma_e B$, where we choose
$B=B_{dark}$, the magnetic field at which we want to study nuclear spin relaxation.
At $t=0$ the pump beam is switched on, and the magnetic field $B_{pump}=180$~G is set.
One can see that the frequency of the SN peak starts to increase. Because the position of the spectral peak in SN
follows the evolution of $B_N$, we  can follow  the build up of the nuclear spin polarization.
As in the case of the FR, the bi-exponential evolution of the peak frequency can be clearly observed in the insulating Sample E, while in the metallic Sample A the evolution is mono-exponential. 
%
%
%
Nuclear spin dynamics under optical pumping  as a function of the magnetic field was previously considered in many works \cite{OpticalOrientation,PagetAmand}.
Here we concentrate on the nuclear spin properties "in the dark".

To study nuclear spin relaxation in the dark, we switch off the pump beam after pumping time $T_{pump}$,
and set the magnetic field to $B_{dark}$,  for which we want to study the relaxation of nuclear spin polarization.
The SN peak starts to move towards $\nu=\gamma_e B_{dark}$, reflecting nuclear depolarization dynamics.
Depending on the electron concentration (metallic or insulating sample) this dynamics is quite different \cite{Ryzhov2015}.
In the insulating Sample E, shown in Fig. \ref{protocol}~(c), the decay is bi-exponential, while in metallic sample, a monoexponential behaviour is observed (see Ref. \onlinecite {Ryzhov2015}).
As has already been noted, this difference can be explained assuming that nuclear spin-lattice relaxation in
 the metallic semiconductor is mediated by itinerant Fermi-edge electrons via the Korringa mechanism,
 while in the dielectric phase it is mediated by donor-bound electrons.
 In the former case, the nuclear spin polarization decays with equal speed at any spatial point.
 In the latter case, the polarization of nuclei under the orbits of donor-bound electrons decays much more rapidly than in the space between donors, where relaxation goes via nuclear spin diffusion towards donors, which play the role of killing centers \cite{Paget82,Giri2013}.
 Such relaxation scenario results in two drastically different decay times for nuclear spin polarization.
 In this paper we focus on studies of spin relaxation of bulk nuclei, not directly affected by contact hyperfine interaction.
 Our goal is to compare the corresponding relaxation times in the samples with different donor densities.
 Thus, from the bi-exponential decay observed in insulating samples we extract the longest decay time, related to the  spin relaxation of the bulk nuclei.
 \subsection{Comparison between SN, PL and FR techniques}
To check the consistency of the results obtained by different techniques, we performed the measurements using both SN and PL techniques in Sample D,
and using both FR and SN techniques in Sample E.
The nuclear spin relaxation times obtained by the different methods on the same sample are identical, within the experimental accuracy.
This is important, because depending on the samples, the measurements were realized by different techniques, as summarized in Table \ref{tab:samples}.
Indeed, the PL experiments were not possible to realize in microcavity samples, while
SN and FR are greatly facilitated by the presence of the cavity.

An important difference between FR and SN experiments is that while the Overhauser-field-induced FR does not require the presence of electrons to detect nuclear
magnetization, the SN signal comes only from regions where resident electrons are present.
Nevertheless, SN signal does provide the information on the spin of the bulk nuclei, due to nuclear spin diffusion from the bulk towards the donor sites.
This is clearly manifested by the presence of the additional fast component in the decay of the spin polarization in the insulating sample observed by SN spectroscopy, while this decay is monoexponential, when FR or PL is measured.
Thus, these methods are complementary and the comparison between them makes it  possible to separate the contributions of
 nuclei with stronger (close to donor sites)  or weaker (bulk nuclei)  coupling to localized electrons in n-type structures.

 The PL experiments are much more time-consuming than SN and FR, because a separate measurement should be realized for
 each duration of the dark interval.
 Another drawback is that it requires injection of the photocarriers for the measurement.  On the other hand, the advantages of the PL include the true "dark" relaxation. 
 Indeed, there is no probe beam, that could, even being $20$~meV below the band gap, have some influence on the nuclear spin dynamics \cite{Ryzhov2016}.  In addition, this technique can be easily applied to thin layers, while the detection of the SN and FR without the cavity enhancing the signal is more demanding.
Finally, the possibility to compare the data via cross-checking procedure involving different methods provides an additional degree of confidence in the obtained results.

%
%

\begin{figure}[]
\center{\includegraphics[width=1\linewidth]{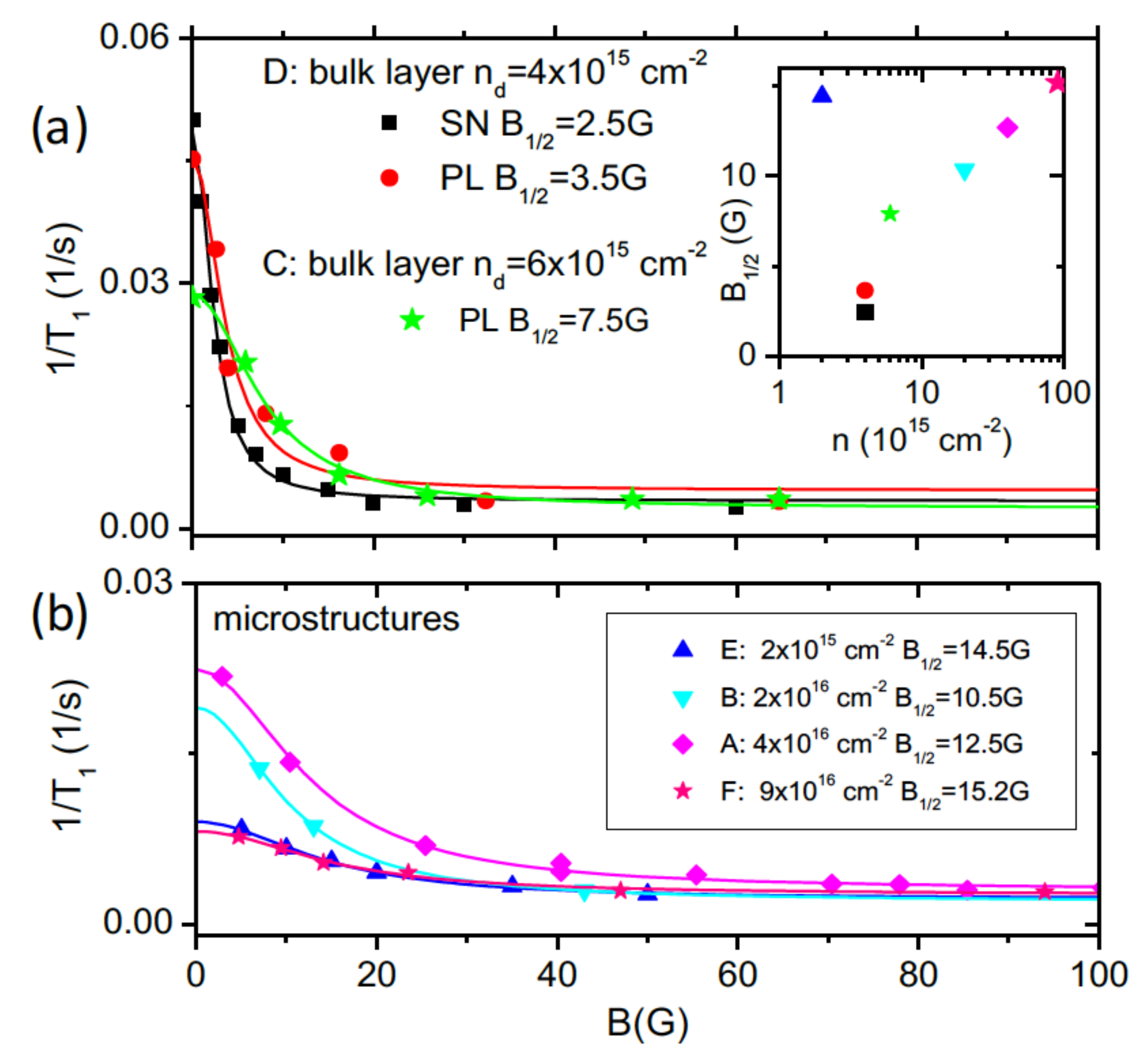} }
\caption{Spin relaxation rate measured as a function of applied magnetic field for different samples at $T\sim4-5$~K.
Symbols are the relaxation rates extracted from the data, solid lines are Lorentzian functions corresponding to the best fit to the data. Bulk samples (a) and microstructures (b) are shown separately. Inset shows the half width at half maximum of the Lorentzians for all Samples.} \label{fig:all_samples}
\end{figure}
\begin{figure}[t]
\center{\includegraphics[width=1.0\linewidth]{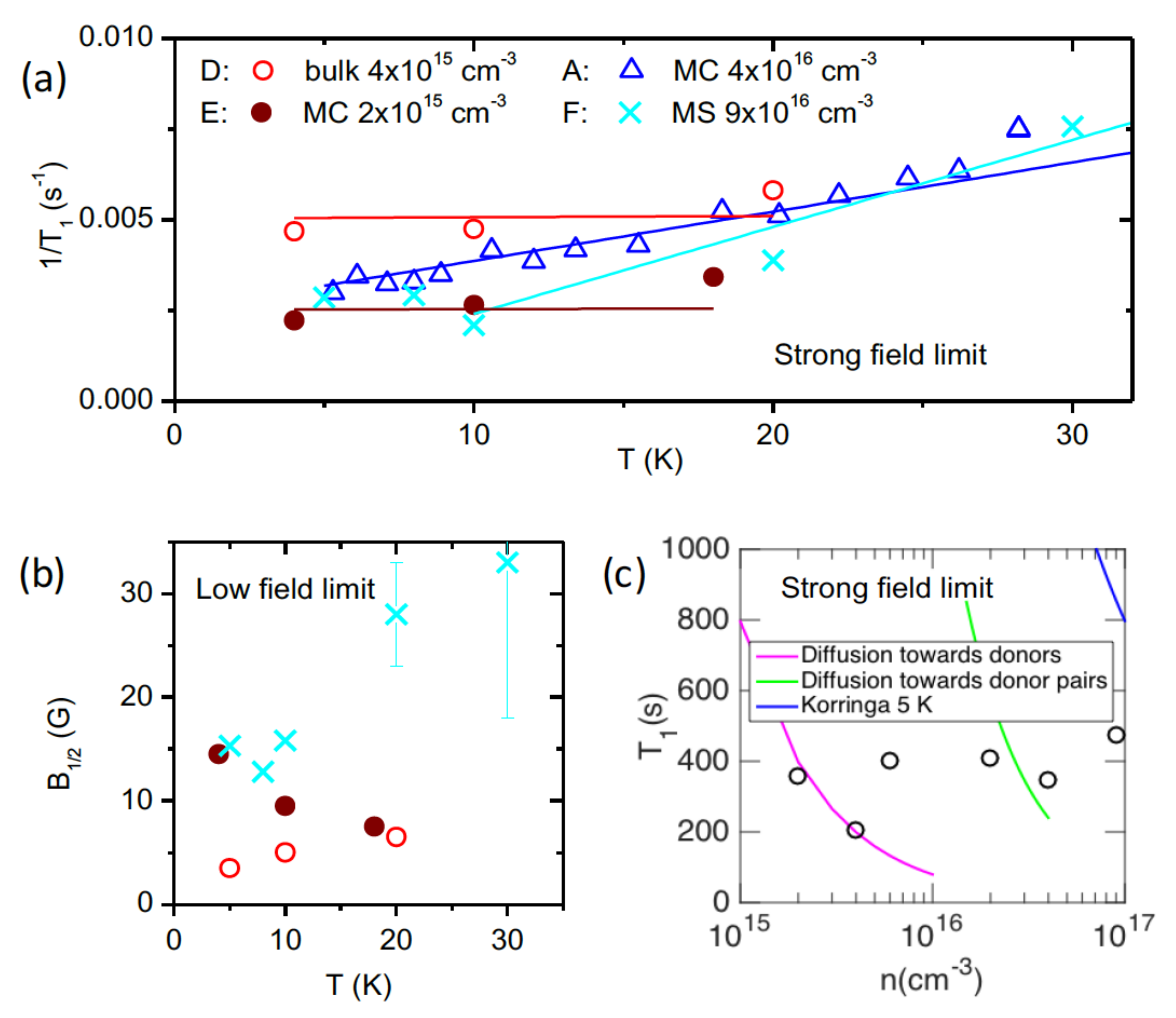} }
 \caption{Temperature dependence. (a) Nuclear spin relaxation rate in the strong field limit, the data (symbols) are obtained from the Lorentzian fits of the measured field dependence at each temperature. Solid lines are  linear temperature dependences expected in metallic samples (Eq. \ref{eq:Tkp}), and a square root dependence expected for the insulating samples (Eq. \ref{eq:TABD}).
 (b) Half width at half maximum of the nuclear relaxation rate  field dependence,  extracted from the Lorentzian fits as those shown in in Fig. \ref{fig:all_samples}. Broadening (narrowing) of the field dependence in different samples. (c) Nuclear spin relaxation times in the strong field limit at $T \sim 5$~K (symbols) for different samples, compared to the theoretical estimations within the diffusion limited relaxation on isolated donors  in insulating samples (Eq.\ref{eq:Td}, magenta line) and on donor pairs  in metallic samples (Eq.\ref{eq:Tdp}, green line). Calculation using Korringa formula (Eq. \ref{eq:korringa}) at $T=5$~K is shown by blue line.
}
\label{fig:korringa}
\end{figure}
\section{Results and discussion}
\label{discussion}
The summary of the nuclear spin relaxation rate measurements
at lowest temperatures as a function of magnetic field is given in Fig. \ref{fig:all_samples}.
Panel (a) shows the two bulk samples, and panel (b) the four microstructures.
%
The salient feature of these data is that the relaxation rate increases dramatically when magnetic field decreases down to zero.
Qualitatively, the behaviour is similar in all samples.
This is the main experimental finding of this work.
Fitting the data to the Lorentzian function, we extract the relaxation rates in the
strong field limit (Fig. \ref{fig:korringa}~(c)) and the
half width at half maximum $B_{1/2}$ of the Lorentzian (Fig.  \ref{fig:all_samples}, inset).
One can see, that the characteristic field $B_{1/2}$,  below which the relaxation rate increases,
is smaller in bulk samples than in the microstructures.
On the other hand  the value of $B_{1/2}$ is not correlated with the donor concentration in the samples.
Indeed, the largest values of $B_{1/2}\sim 15$~G are observed in the samples with the lowest and  the 
highest donor densities (Samples E, F).

The behaviour of the relaxation times in the strong field limit is not less surprising. In the range of the
studied donor densities, no significant variation of the nuclear spin relaxation is observed (Fig. \ref{fig:korringa}~(c)).

To  complete the analysis, we explore the temperature dependence of the nuclear spin relaxation.
The strong field limit of the nuclear spin relaxation is shown in Fig. \ref{fig:korringa}~(a).
It was measured under magnetic field ranging from $150$ to $1000$~G, where no field dependence is observed.
%
In the insulating Samples C, D and E we do not observe any pronounced effect of the temperature  (at least up to $20$~K),
while in metallic Samples A and F nuclear spin relaxation rate increases with temperature.
The  values of $B_{1/2}$ obtained  from the fits to the Lorentzian shape of the relaxation rate
as a function of the magnetic field are shown in Fig.  \ref{fig:korringa}~(b).
One can see, that there is no systematic behaviour,  so that we can't associate  it either
with the donor density or with the presence of the microstructure.

In the following we discuss possible mechanisms of nuclear spin relaxation that would allow for self-consistent description of these data.
An analysis of  the strong field relaxation is followed by a discussion of the magnetic field dependence.
\subsection{Strong field nuclear spin relaxation.}
In order to understand these experimental results, we start from the strong field and low temperature regime.
Under these conditions nuclear spin relaxation rate is only weakly affected by the density of donors.
In Fig.  \ref{fig:korringa}~(c) we compare the data with the theoretical estimations.
At lowest donor densities, nuclear spin dynamics is controlled by the fast spin relaxation due to hyperfine coupling in the vicinity of the donor sites, and by the diffusion of the nuclear spin toward the donor sites. In this case the
relaxation rate is given by Eq.~(\ref{eq:Td}).
Calculation using this formula without any fitting parameters describes quite well the data at lowest density (Samples D, E).

We have shown in our previous work \cite{Giri2013}, that close to the MIT, despite the presence of the delocalized electrons on the Fermi level, the nuclear spin relaxation is still dominated by the nuclear spin diffusion towards the "killing centers", where efficient relaxation via hyperfine interaction with electrons takes place.
The role of the "killing" centres" is played in this case by donor pairs, that can still localize an electron despite the presence of the electron gas.
The spatial distribution of the donors in the sample can be supposed to obey the Poisson distribution.
Under this assumption, for the density of the  donor pairs separated by less than the screening length of the Coulomb potential in the electron gas, one can obtain the following expression:
\begin{equation}
n_p=p n_d^{3/2}\exp(-p n_d^{1/2}),
\end{equation}
where
$p=\frac{\pi}{6}(\frac{\pi}{3})^{1/2} a^{3/2}$.
Then, the diffusion-limited relaxation rate is given by Eq.~(\ref{eq:Td}), with the donor density replaced by the density of donor pairs,
and the  donor localization radius $a$ by the localization radius of the donor pair $a_p$:
 \begin{equation}
 T_{p}^{-1}\approx 4 \pi D n_p a_p.
 \label{eq:Tdp}
 \end{equation}
Assuming random distribution of the donor positions, it is reasonable to 
estimate the localization radius of the donor pair as the size of largest pair that can localize an electron. 
It is given by the screening length of the electron Fermi gas: 
\begin{equation}
a_p=\frac{1}{2}\left(  \frac{\pi}{3} \right)^{1/6}   \left(  \frac{a}{n^{1/3}} \right)^{1/2}.
\end{equation}
Here $n$ is the electron gas density, which we assume equal to $n_d$.
The numerical application of this formula is shown in Fig.~\ref{fig:korringa}~(c) by the green line.
One can see that it provides a satisfactory description of the nuclear relaxation time for Samples A and B, characterized by the
density of donors on the metallic side of the MIT.

Sample C falls in the intermediate regime, where the donor density $n_d	\simeq 0.5 n_c$, $n_c$ being the critical density for the MIT.
Neither isolated donor (which underestimates), nor donor pair model (which overestimates) can give the correct value of the nuclear spin relaxation time.
In the sample with the similar donor density $n_d=5.9\times10^{15}$~cm$^{-3}$ Lu et al  \cite{Lu} have measured $T_1=1250$~s at $1.55$~T. 
This value is $3$ times higher than our result, and it does not fit the diffusion model either.
 Lu et al explained this long relaxation time (compared with diffusion limited model prediction) 
 by the diffusion barrier, which builds up around the donors, due to inhomogeneous Knight field \cite{Paget82}.
 This model does not seem applicable to our experimental conditions, because in our case the applied field is $100$ times smaller, 
 and so is the resulting electron spin polarization and the Knight field.
 More detailed comparison with Ref. \onlinecite{Lu} requires application of the strong magnetic fields, which was not possible in this work.

In the most heavily doped Sample F, the relaxation time substantially exceeds the prediction of the diffusion limited relaxation model  (diffusion towards donor pairs in this case).
This can be explained by the reduced number of the donor pairs that can localize  an electron, due to the efficient screening of the attractive potential of
donor pairs by the free electrons.
However, the relaxation time measured in Sample F at low temperature ( $T=5$~K ) is still shorter than the time predicted by the Korringa formula \cite{Korringa1950}:
\begin{equation}
T_K^{-1}=\frac{\pi}{\hbar}A^2 \nu_0^2 \rho^2(E_F)k_BT.
\label{eq:korringa}
\end{equation}
Here $A=\frac{8 \pi}{3}\gamma_e \gamma_N \hbar^2$, $\gamma_N$ is nuclear gyromagnetic ratio.
We use the average value $A=44$~$\mu$eV between $A_{As}=46$~$\mu$eV and  $A_{Ga}=42$~$\mu$eV.
 $\nu_0=0.044$~nm$^3$ is the primitive cell volume, and $\rho(E_F)$ is the density of states at the Fermi level $E_F$.
 The result of this calculation for two different temperatures is shown in Fig. \ref{fig:korringa}~(c) for $T=5$~K (blue line).

To get deeper  insight in the role of Korringa mechanism in the nuclear spin relaxation,
we compare the measured temperature dependence of the nuclear spin relaxation rate in
the strong field limit ($B>100$~G) to the linear dependence expected from Eq. \ref{eq:korringa}.
 For Sample A  we do observe linear increase of the relaxation rate with temperature from $5$ to $30$~K (blue triangles in Fig.  \ref{fig:korringa}~(a)).
 However, to fit the experimental data, a linear dependence with the slope given by Eq. \ref{eq:korringa} must be shifted
 by a constant value, corresponding to the relaxation rate measured at  $5$~K, and interpreted as the relaxation limited by the diffusion towards the donor pairs  $T_K^{-1}$ (blue line).
 This suggests that  Korringa mechanism dominates   nuclear spin relaxation at $T>5$~K,
 and the total spin relaxation rate is given by
 \begin{equation}
  T_1^{-1}=  T_K^{-1}+  T_p^{-1}
  \label{eq:Tkp}
 \end{equation}
The picture is slightly different for Sample F (cyan crosses in Fig.~\ref{fig:korringa}~(a)).
The linear increase of the nuclear spin relaxation rate with temperature is only observed above $10$~K.
The slope fits perfectly the result of Eq.~(\ref{eq:korringa}). In contrast with Sample A, no offset
related to the spin diffusion needs to be assumed.
This could mean that the number of  donor clusters localizing an electron is strongly temperature dependent in this heavily-doped sample.
At low temperatures there exist a certain number of localized electrons that contribute to the diffusion-limited hyperfine relaxation of the nuclear spin, while at higher temperatures the electrons get delocalized, resulting in a  purely Korringa-type relaxation.
Note, that this argument  agrees perfectly with the previous results, demonstrating that nuclear spin relaxation times at temperatures from $10$ to $30$~K and under magnetic fields up to $1200$~G are described by the Korringa formula Eq. (\ref{eq:korringa}) without any fitting parameters \cite{KoturJETP2014}.

%
%

For the insulating samples the temperature dependence is expected to be weak, and we do not observe any noticeable temperature dependence (Fig.~\ref{fig:korringa}~(a)).
Indeed, neither nuclear spin diffusion, nor hyperfine relaxation are temperature dependent.
The only contribution comes from the relaxation by the spin fluctuations of the free electrons.
The relaxation rate of the nuclei due to interaction with non-degenerate gas of free electrons
is given by the  Bloembergen-Abragam formula \cite{Bloembergen1954,Abragam,OpticalOrientation}:
\begin{equation}
T_{BA}^{-1}\sim \frac{2}{(2 \pi)^{3/2}} A^2  \nu_0^2 m_e^{3/2} \hbar^{-4} (k_B T)^{1/2}
\label{eq:BA}
\end{equation}
This formula dramatically underestimates the observed relaxation rate at all the temperatures that could be accessed in this work, so that the total relaxation rate
 \begin{equation}
  T_1^{-1}=  T_{AB}^{-1}+  T_D^{-1}
  \label{eq:TABD}
 \end{equation}
is largely dominated by diffusion towards donors sites localizing  an electron.

Thus, we obtain the following picture for the strong field relaxation in n-GaAs:
At low donor density $n_D < 5 \times 10^{15}$~cm$^{-3}$ the relaxation of bulk nuclei is limited by the diffusion towards donors, where fast relaxation by hyperfine interaction takes place. The relaxation rate increases with increasing donor density.
At higher densities, but still below MIT  $5 \times 10^{15}$~cm$^{-3} < n_D < 2 \times 10^{16}$~cm$^{-3}$,  the bulk nuclei start to be situated on the outer shell of the localized elections. In this case the  diffusion-limited  mechanism does not describe the relaxation correctly, and neither does the direct hyperfine coupling.
Above MIT, localization potential of the single donor is screened, but the pairs of closely lying
donors can still localize an electron, at least at low temperatures.
Therefore, at $1 \times 10^{16}$~cm$^{-3} < n_D < 5 \times 10^{16}$~cm$^{-3}$ the spin relaxation is limited by diffusion towards the donor pairs, while at higher temperatures ($T>5$~K) the relaxation due to the electron spin fluctuations at the Fermi level becomes dominant (Korringa relaxation).
At even higher densities  $n_D > 5 \times 10^{16}$~cm$^{-3}$, electron gas starts to screen the attractive potential of the donor pairs, so that they can only localize electrons at low temperatures.
%
%
Above $10$~K the  Korringa relaxation remains the only nuclear spin relaxation mechanism.
\subsection{Magnetic field dependence of nuclear spin relaxation.}
The magnetic field dependence of the nuclear spin relaxation rate shown in Fig. \ref{fig:all_samples} is another important result of this work.
Let us recall, that at low magnetic fields, comparable to  local field $B_L$ (given by dipole-dipole interactions within the nuclear spin system)  the non-equilibrium nuclear angular momentum decays within the spin-spin relaxation time $T_2$ of order of $100$~$\mu$s.
Because the characteristic time of the energy transfer between the nuclear spin system and the crystal lattice is many orders of magnitude longer than $T_2$, a partial equilibrium  establishes in the nuclear spin system within $T_2$.
It is characterized by the nuclear spin temperature $\Theta_N$. 
Thus, in the presence of the external magnetic field the polarization of the nuclear spin system is induced via its paramagnetic susceptibility, which  is inversely proportional to $\Theta_N$.
The latter relaxes towards the lattice temperature $T$ with the  spin-lattice relaxation time $T_1>>T_2$.
For this reason, the nuclear spin relaxation at low field is rather the warm up of the nuclear spin system, which is determined by energy transfer between the nuclei and the crystal lattice.

None of the nuclear spin relaxation mechanisms discussed above (hyperfine interaction, spin diffusion, coupling to free electrons) can account for the enhancement of the nuclear spin warm up at low field.
In particular at the low magnetic fields considered here, a spin diffusion barrier, susceptible to slow down the nuclear spin relaxation, is not expected to form \cite{Paget82, Lu}.

We have recently suggested, that in insulating samples this effect can result from the interaction of nuclear quadrupole moments with electric field gradients induced by slow spatiotemporal fluctuations of localized electron charges, provided that the corresponding
correlation time $\tau_c^c>>T_2$ \cite{Kotur2016}.
This theory shows  that the energy flux between nuclear spin and electron charge via slowly varying quadrupole interaction ${F}_Q$ does not depend on the magnetic field, while  the heat capacity of the nuclear spin system $C_N$ is strongly field dependent.
The corresponding field-dependent relaxation time
\begin{equation}
T_Q^{-1}=\Theta_N F_Q C_N^{-1}
\label{eq:tq}
\end{equation}
provides an additional contribution to the total relaxation rate in the insulating samples:
\begin{equation}
T_1^{-1}=T_D^{-1}+T_{AB}^{-1}+T_Q^{-1}.
\label{eq:tqd}
\end{equation}
The quadrupole relaxation rate $T_Q^{-1}$ vanishes at $B>>B_L$, but can be important at
low magnetic field \cite{Kotur2016}:
\begin{equation}
T_Q^{-1}=\frac{4 \pi \mathcal{L}   (eQ \beta_Q E_a)^2}{5 (\hbar \gamma_N)^2 (B^2+B_{L}^2)\tau_c^c}
\frac{4I(I+1)-3}{(8I(2I-1))^2}
\label{T_Q}
\end{equation}
Here $I$ is the nuclear spin,   $\mathcal{L}$ is the dimensionless coefficient that accounts for the averaging of the electric fields from the electrons, $\beta_{Q}$ is the experimentally determined and isotope-dependent constant, $eQ$ is  the nuclear quadrupole moment, also isotope-dependent, $e$ is the electron charge, $E_a$ is the electric field at Bohr radius distance  from the charged donor positon \cite{Harris2002}.

In Ref. \onlinecite{Kotur2016} we have successfully applied the above ideas to
interpret the data obtained in Sample D by PL experiments.
Here we first of all confirm the experimental results of Ref. \onlinecite{Kotur2016}
by an alternative experimental technique of SN.
Such comparison confirms an estimation for the precision of the $B_{1/2}$ measurements for this
sample $\delta B_{1/2}  \sim 2$~G that was given in Ref. \onlinecite{Kotur2016}.
Note, that $B_{1/2}$ should be  interpreted in this model as the local field $B_L$,
characterizing various interactions within nuclear spin system, while the height of the Lorentzian is $1/T_1^{(B=0)}-1/T_1^{(B=\infty)}$.

The application of the quadrupole relaxation theory to other insulating samples could be straightforward.
However, the value of the $B_{1/2} \sim 15$~G in Sample E is much higher than values that could \textit{a priori} be expected for the local field.
Indeed the most well-known contribution results from spin-spin interactions
$B_{SS}=1.5$~G \cite{Paget1976}.
The missing part of the local field could be attributed to the quadrupole interactions or to some spin-spin interactions not accounted for in Ref. \onlinecite{Paget1976}, such as Dzyaloshinskii- Morya indirect exchange interaction.
But the most plausible explanation is that the large values of the local field result
from the strain-induced quadrupole splitting in the microstructure samples.
Indeed, the  difference in $B_{1/2}$ between bulk samples and various microstructures
is so important that they are presented in the two different panels in Fig. \ref{fig:all_samples}.
Similar effects of the strain are well known in semiconductor quantum dots, where
strain-induced  quadrupole splittings are so large, that even  the concept of the spin temperature can not be applied any more \cite{MaletinskyNatPhys2009}.
The observed temperature dependence of the $B_{1/2}$ in the insulating samples is consistent
with this interpretation (Fig. \ref{fig:korringa} (b)).
In the bulk Sample D  $B_{1/2}$ is not substantially affected by temperature, while in the microcavity Sample E it decreases with temperature, because of the strain relief.
Thus, in the insulating samples we attribute the enhancement of the nuclear warm up rate at low field  to the quadrupole-induced mechanism of spin relaxation.
The characteristic field for this enhancement is strongly strain-dependent, and is strongly increased in microstructures with respect to bulk samples.
In metallic samples the enhancement of  the nuclear warm up rate at low field by a factor of $\sim 3$ can, 
in principle be expected within Korringa relaxation mechanism. 
Indeed, since the Fermi length of the electron gas is much larger than the lattice constant, the 
fluctuating hyperfine fields created by the electrons and acting on the nuclei are strongly correlated.
This fact was taken into account in the calculations by Abragam, predicting the enhancement factor $\xi=3$
at $B<<B_L$, see Eq. IX.20 in Ref. \onlinecite{Abragam}. 
However, this calculation is only valid when the local field is governed by the dipole-dipole interactions $B_L=B_{SS}\sim 2$~G, and thus can not be directly applied to our results.
Generalisation of this approach to the quadrupole interaction yields $\xi \sim 1$ \cite{Wolf},
which means no enhancement of the spin relaxation at $B<<B_L$.
Thus, to understand the low-field enhancement of the nuclear spin relaxation in metallic samples,
one needs to  search for a suitable source of the slowly changing fluctuations.
Indeed, long-range electric fields are efficiently screened by the free electron gas.
Moreover, the model developed for the insulating samples requires $\tau_c^c>>T_2=100$~$\mu$s,
which is not so obvious for the metallic samples.
In addition, in Sample F at $T>10$~K the value of $B_{1/2}$ increases dramatically.
The underling mechanism is yet to be identified.

%
%
%
%
\section{Conclusions}
\label{Conclusions}
We have studied nuclear spin relaxation in the set of n-GaAs samples with donor concentrations
varying across MIT from insulating ($n_d=2\times10^{15}$~cm$^{-3}$) to metallic regime  ($n_d=9\times10^{16}$~cm$^{-3}$).
Three different experimental techniques were applied that provided consistent results: PL with dark intervals, spin noise spectroscopy, and Faraday rotation.
All these methods allow to study nuclear spin relaxation in the absence of the photocreated carriers.

Under magnetic field $B \ge 100$~G, we identified  different regimes of spin relaxation.
At low donor density $n_D < 5 \times 10^{15}$~cm$^{-3}$ the relaxation of bulk nuclei is limited by the diffusion towards donors, where fast relaxation by hyperfine interaction takes place.
At higher densities, but still below MIT  $5 \times 10^{15}$~cm$^{-3} < n_D < 1 \times 10^{16}$~cm$^{-3}$ the nuclear spin lifetimes are longer than predicted by the diffusion-limited hyperfine relaxation model. This result is consistent with previous findings but has no plausible theoretical explanation so far.
%
%
Above MIT, the localization potential of the single donor is screened, but the pairs of the closely lying
donors can still localize an electron, at least at low temperatures.
Therefore, at $1 \times 10^{16}$~cm$^{-3} < n_D < 5 \times 10^{16}$~cm$^{-3}$ the spin relaxation is limited by diffusion towards the donor pairs.
At higher temperatures ($T>5$~K) the relaxation due to the electron spin fluctuations at the Fermi level becomes dominant (Korringa relaxation).

At low magnetic field we found that nuclear spin relaxation rate increases for all the samples.
Such behavior suggests that the relaxation is caused by  slowly fluctuating fields (either electric or magnetic), characterized by the correlation times $\tau_c>T_2$.
On the insulating side of the MIT this effect can be understood as a result of the interaction of the quadrupole moment of the nuclei with slowly fluctuating electric fields, due to hopping of the electron charge, either into conduction band, or across the impurity band.
On the metallic side of the MIT the possible origin of the slowly fluctuating fields should yet be identified.

\section*{Acknowledgements} This work was supported 
 by the joint grant  of the Russian Foundation for Basic Research (RFBR, Grant No. 16-52-150008)   and
 National Center for Scientific Research (CNRS, PRC SPINCOOL No. 148362), as well as French National Research Agency (Grant  OBELIX, No. ANR-15-CE30-0020-02).
 RD and MK acknowledge the support  of the Ministry of Education and Science of the Russian Federation (contract
No. 14.Z50.31.0021 with the Ioffe  Institute, Russian Academy of Sciences,
and leading researcher M. Bayer). VSZ acknowledges the support of the RFBR (grant No. 15-52-12013).

%

\end{document}